# Observation of Zeeman shift in the rubidium D$_2$ line using an optical nanofiber in vapor

Amy Watkins[a], Vibhuti Bhushan Tiwari[b], Jonathan M. Ward[c], Kieran Deasy[c] and Síle Nic Chormaic[*,a,c,d]

[a]Physics Department, University College Cork, Cork, Ireland; [b]Laser Physics Applications Section, Centre for Advanced Technology, Indore 452013, India; [c]Light-Matter Interactions Unit, OIST Graduate University, Onna-son, Okinawa 904-0495, Japan; [d]School of Chemistry and Physics, University of Kwa-Zulu Natal, Durban 4001, South Africa

## ABSTRACT

We report on the observation of a Zeeman shift (order of 100 MHz) of the Doppler-broadened D$_2$ transition of both $^{85}$Rb and $^{87}$Rb isotopes via transmission through a 400 nm diameter optical nanofiber in the presence of a DC magnetic field. Linearly-polarized light propagating in the nanofiber is analyzed as a superposition of two orthogonally circularly polarized orientations, $\sigma^+$ and $\sigma^-$. In the absence of the magnetic field, the absorption of these polarizations by the atomic vapor, via the evanescent field at the waist of the nanofiber, is degenerate. When a weak magnetic field is applied parallel to the propagating light, this degeneracy is lifted and relative shifts in the resonance frequencies are detected. Typical linear shift rates of 1.6 MHz/G and -2.0 MHz/G were observed. We also demonstrate a dichroic atomic vapor laser lock line shape by monitoring the real-time subtraction of the two magnetically-shifted absorption spectra. This is particularly interesting for magneto-optical experiments as it could be directly implemented for diode laser frequency-stabilization.

## 1. INTRODUCTION

Tapered optical fibers[1] have been used extensively in a range of diverse applications from atom-optics tools[2-5], through highly-efficient optical coupling devices[6,7] to spectroscopy[8,9]. They consist of a subwavelength-diameter waist region, which adiabatically expands to standard optical fiber. In each application, the evanescent field created at the waist provides a means for photon interactions with the surrounding environment. More specifically one can define the nanofiber as being active, i.e., a probe beam propagates through it[2,5,10] or passive, i.e., there is no probe beam present[3,4]. In the active case, a low input power (~nW in typical experiments) results in a high intensity at the waist due to the small mode area. This facilitates interactions with, for example, an atomic vapor surrounding the fiber and the realization of Doppler-broadened and sub-Doppler absorption spectroscopy[2,8], two-photon absorption[5], and low power modulation due to the quantum Zeno effect[11].

Here, we report on a novel technique to use the absorption at the nanofiber waist to monitor the Zeeman shift of the D$_2$ lines in Rb in the presence of a weak magnetic field. In conventional free-space magneto-optical techniques[12] a linearly polarized probe beam incident on a vapor cell is seen by the atoms as a superposition of $\sigma^+$ and $\sigma^-$ circularly polarized orientations. In the absence of a magnetic field, the two atomic transitions are degenerate and absorb both polarizations equally. On the other hand, in the presence of a DC magnetic field, this degeneracy is lifted (the Zeeman shift) and the system becomes dichroic. As a result, the $\sigma^+$ and $\sigma^-$ spectra are absorbed at higher and lower frequencies, respectively and the resultant frequency shift is given by[13]

$$\delta\omega = g_F \frac{\mu_B B}{\hbar} m_F, \qquad (1)$$

where $g_F$ is the Landé g-factor, $\mu_B$ is the Bohr magneton, $B$ is the applied magnetic field and $m_F$ is the magnetic quantum number. Consequently, the subtraction of these shifted spectra results in a wide-capture range differential signal coined as a dichroic atomic vapor laser lock (DAVLL) line shape[14]. This signal can then be used for laser frequency stabilization by locking to the dispersive-like gradient around one of the resonance frequencies[14,15].

---

[*] sile.nicchormaic@oist.jp

In this work, we use an optical nanofiber as an effective substitute for the traditional free-space probe beam to measure the corresponding Zeeman shift and a preliminary DAVLL line shape. Initially, a linearly-polarized beam is coupled into the nanofiber installed in a vapor with both $^{85}$Rb and $^{87}$Rb isotopes present. The transmitted power out the end of the fiber is passed through a circular analyzer setup[16], consisting of a λ/4 plate and a polarizing beam splitter (PBS), to separate and monitor the σ$^+$ and σ$^-$ polarizations. When a magnetic field of 60 G is applied to the system the Zeeman shift of the two orthogonally-polarized orientations about the center absorption frequency is observed. We also present a DAVLL spectrum from the subtraction of the oppositely-shifted spectra detected from the nanofiber transmission.

## 2. EXPERIMENTAL DETAILS AND RESULTS

The nanofiber is fabricated from 780 nm single-mode optical fiber (Newport FS-E) using a direct heat-and-pull technique with an oxy-butane flame[3]. After fabrication, the tapered fiber is mounted on an aluminum support using UV-epoxy and installed in a vacuum chamber (with a base pressure of $1\times10^{-5}$ mbar) using a standard Teflon fiber feed-through system[3]. A fiber with a waist of approximately 400 nm and an optical transmission of 90% prior to mounting is used. During the experiment, the chamber is heated to 50ºC and Rb vapor is distributed in the chamber by resistively heating a getter dispenser (SAES Getters). A 200 nW probe beam from a 780 nm diode laser (Toptica DL100) is coupled into the fiber with a coupling efficiency of ~75%. A λ/2 plate placed before the coupling system controls the power and purity of the linearly-polarized light launched into the fiber. The transmitted power is decoupled out the other end of the fiber where a λ/4 plate circularly polarizes the beam and a PBS, orientated at 45º to the fast axis of the λ/4 plate, separates the σ$^+$ and σ$^-$ components onto two avalanche photodiode detectors (APDs). A schematic of the experimental setup is shown in Fig. 1.

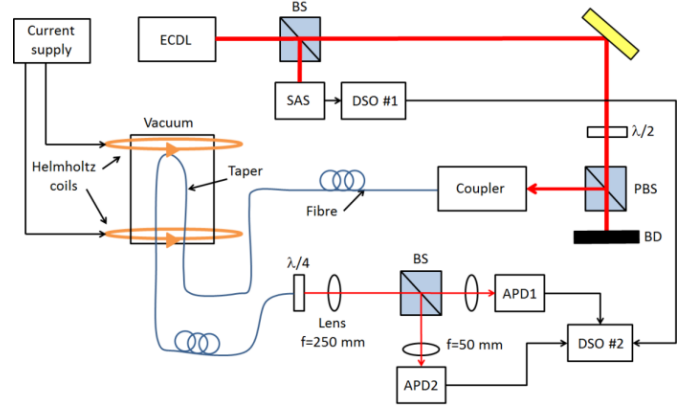

Fig. 1 (Color online) Schematic of experimental setup. ECDL: extended cavity diode laser, BS: beam splitter, SAS: saturated absorption spectroscopy, DSO: digital storage oscilloscope, PBS: polarizing beam splitter, BD: beam dump, APD: avalanche photodiode.

The laser frequency is scanned over the sub-Doppler spectrum of the $D_2$ transition of Rb using a saturated absorption spectroscopy (SAS) signal. The corresponding spectrum is recorded using a Si photodiode (PD) and a two-channel digital signal oscilloscope (DSO) (see top trace in Fig. 2). As Rb vapor is added to the chamber, the σ$^+$ and σ$^-$ nanofiber signals on the APDs (see Fig.1) are monitored and averaged 64 times via a second two-channel DSO, triggered from the linear ramp of the laser frequency scan. In the absence of a B-field, the reference spectrum and the two nanofiber spectra are recorded simultaneously, see the solid line traces in Fig. 2. The data clearly show a red-shift in the range of 50-125 MHz with respect to the free-space signals, depending on the transition. This is primarily attributed to collisional and time-transit broadening, with a possible small contribution arising from the residual magnetic field (order of 4 G equivalent to a shift of approximately 5.6 MHz) from the heating strap on the vacuum chamber. In line with convention, in the absence of a magnetic field, the circularly polarized components shown in Fig. 2 are degenerately absorbed by the Rb vapor at approximately the same resonance frequencies. From the absorption signals, we estimate that the number of atoms interacting with the evanescent field of the tapered fiber to be around 500 with a vapor density of $1\times10^{12}$ atoms/cm$^3$.

Next, a pair of Helmholtz coils, each with an inner diameter of 7 cm and 60 turns, is incorporated into the setup so that the resulting magnetic field is parallel to the direction of light propagation in the nanofiber. The coils are separated by a distance of 3.8 cm and they are arranged so that their centers are aligned with the waist of the nanofiber. A current of 5 A is applied to the coils, generating a field of 60 G, and the $\sigma^+$ and $\sigma^-$ nanofiber spectra are recorded as before (see the dashed lines in Fig. 2). It was necessary to slightly rescale the intensities of the shifted spectra (B-field present) to the unshifted spectra (B-field absent) for clarity to observe the magnetically-induced shift. Again, in accordance with convention, in the presence of the magnetic field the degeneracy of the two circularly-polarized orientations is lifted. The absorption spectrum is shifted to higher frequencies (blue-shifted) for the $\sigma^+$ component and to lower frequencies (red-shifted) for the $\sigma^-$ component, in other words a Zeeman shift is observed with respect to the signals recorded without an applied magnetic field.

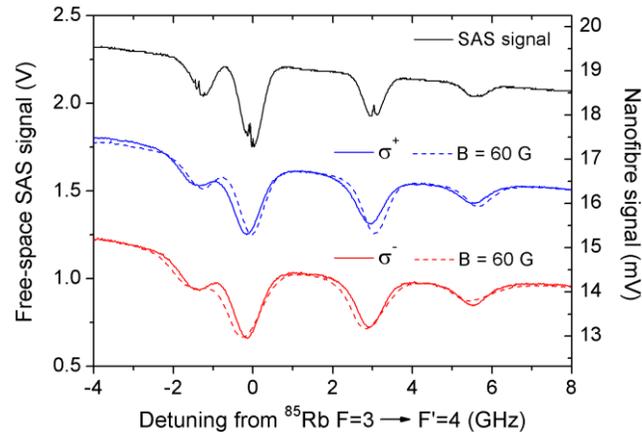

Fig. 2 (Color online) Upper trace: Free space SAS signal; Middle and lower traces: Absorption spectra through the nanofiber (dashed lines) showing the Zeeman shift of the $\sigma^+$ (middle) and $\sigma^-$ polarizations (lower) in the presence of a 60 G magnetic. The fiber signals in the absence of an applied magnetic field are also shown for comparison (solid lines). The dashed curves have been rescaled along the *y*-axis with respect to the unshifted nanofiber spectra (solid lines) for clarity.

To investigate the Zeeman shift rate, the $\sigma^+$ and $\sigma^-$ nanofiber spectra are recorded while increasing current in the magnetic coils in steps of 0.5 A. For each trace, the fiber spectrum is differentiated and the detunings of zero-crossing frequencies of the Doppler-broadened dips are plotted with respect to their free-space counterparts in the SAS spectrum (see Fig. 3). The Zeeman shift of the $^{87}$Rb F=2 to F' transition is omitted due to the presence of time-transit broadening[2], which increases the overlap with the $^{85}$Rb F=3 to F' transition, primarily due to the close proximity of their atomic energies. As a consequence, it is not trivial to ascertain the zero-crossing frequencies of this particular transition with reasonable certainty. Linear fits are applied to each pair of Zeeman shifts for the other three Doppler-broadened transitions: $^{85}$Rb F=3 to F', $^{85}$Rb F=2 to F' and $^{87}$Rb F=1 to F'. The upper and lower data sets in each pair correspond to the $\sigma^+$ and $\sigma^-$ orientations, respectively. One can see that the $\sigma^+$ signal appears to be blue-shifted with respect to estimates of the frequency shift using Eq. 1, whereas the $\sigma^-$ signal appears to be red-shifted. The extent of the red or blue shift depends on the magnitude of applied B-field. Note that there appear to be oscillations in the frequency shift as the magnetic field increases and further studies are being conducted to determine their significance. In addition, the shift rates for each polarization pair in Fig. 3 are not symmetric about the unshifted center frequencies with respect to the theory estimates. This is clearly evident between the $\sigma^+$ and $\sigma^-$ polarization states in Fig. 3(middle), for example. It is important to note that any frequency shift effects arising from atom-surface interactions at the fiber waist[17,18], such as the van der Waals (vdW) effect, were not taken into account in estimating shifts from Eq. 1 and could account for the discrepancies observed here, though our estimates indicate that such shifts would be far smaller (order of 4 MHz)[19] than those observed.

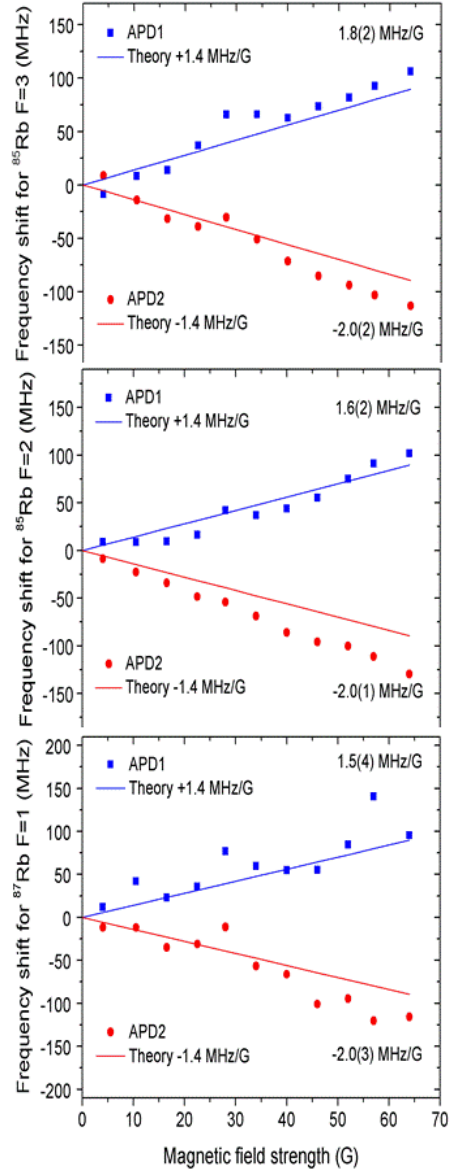

Fig. 3 (Color online) Zeeman shifts of the $\sigma^+$ and $\sigma^-$ orientations for the three Doppler-broadened $D_2$ transitions (top) $^{85}$Rb F=3 to F', (middle) $^{85}$Rb F=2 to F' and (bottom) $^{87}$Rb F=1 to F'. The upper (blue) and lower (red) data in each pair correspond to the $\sigma^+$ and $\sigma^-$ polarizations, respectively. The solid lines represent the estimated shifts from Eq. 1.

In general, as seen in Fig. 3, the observed red-shift for the $\sigma^-$ signal appears to be larger than the blue-shift for the $\sigma^+$ signal. Note that this observed behavior is similar to the experimental red-shift of the atomic resonance when the fluorescence excitation spectrum for Cs atoms distributed around a nanofiber was measured through a nanofiber[18], an effect that was attributed to the van der Waals interaction. However, in our setup, a detailed analysis of optical pumping and fiber-related birefringence effects, including a full analysis of the polarization states in the nanofiber, would be required in order to verify the final source of the shift asymmetry and such a treatment is deemed beyond the scope of this paper, but is the focus of future work.

Finally, the real-time subtraction of the signal transmissions through the nanofiber of the two oppositely circularly-polarized spectra in the presence of a magnetic field results in a dispersive DAVLL line shape (see lower trace in Fig. 4). The absorption levels of the two orthogonal signals were equalized by rotating the $\lambda/2$ plate before the fiber coupler in the absence of the magnetic field. Next a current was applied to the magnetic coils to produce a field of ~60 G, yielding the dispersive-like signal observed, with a range of approximately ±110 MHz about each of the unshifted center frequencies. A maximum gradient of 1.65(4) mV/GHz was recorded for the $^{85}$Rb F=3 to F' transition. The overall shape of the signal and the frequency-crossing points can be altered by a small rotation (of a few degrees) of the λ/4 plate with respect to the PBS. As a consequence of out-coupling the signal from the fiber for the circular polarization analyzer setup, the vertical axis of the DAVLL signal in Fig. 4 is not zeroed. The intensities of the two circularly-polarized spectra are not of equal magnitude when the magnetic field is on and this is simply treated as a residual offset. Note that the low signal levels and the DAVLL slopes could be dramatically improved upon, by realizing an all-fiber polarization analyzer arrangement. Another viable solution for improving the magnitude of the slope would be to incorporate a precision, high-gain, differential amplifier circuit as is routinely employed in conventional free-space setups[14,16]. To use this line shape for laser frequency stabilization, the capture range should be widened to ±500 MHz for an optimum Doppler-broadened locking signal[14]. This can be achieved by incorporating a solenoid to achieve a magnetic field strength of ≥100 G. During the measurements presented here, the intensities of the two spectra had a tendency to vary in magnitude as a function of time. In order to maintain the signal, it was necessary to rotate the λ/2 plate by a few degrees. Thus, we propose that any fiber birefringence present needs to be addressed by implementing a Berek compensator before the fiber coupler[10] or investigating the use of polarization-maintaining tapered fiber[20].

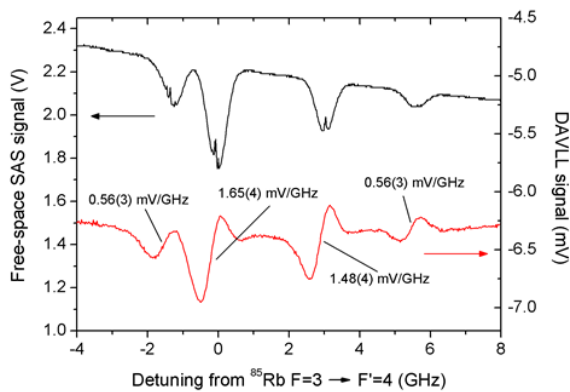

Fig. 4 (Color online) Bottom trace: Nanofiber DAVLL line shape about all four Doppler-broadened resonances, the $^{87}$Rb F=2 to F', $^{85}$Rb F=3 to F', $^{85}$Rb F=2 to F' and $^{87}$Rb F=1 to F' (from left to right) with a magnetic field of 60 G. Upper trace: The SAS signal.

## 3. CONCLUSION

We have demonstrated the ability to measure magnetically-induced shifts of the Doppler-broadened $D_2$ absorption lines of atomic Rb by monitoring the optical transmission through a tapered nanofiber installed in a vapor. The Zeeman shift rates observed for each transition are comparable to the Bohr magneton and provide a method for studying atom-surface interactions. The real-time subtraction of the two circularly-polarized spectra results in a dispersive DAVLL line shape with a maximum gradient of 1.65(4) mV/GHz around the $^{85}$Rb F=3 to F' transition for a magnetic field of 60 G. These preliminary results reinforce the use of a tapered fiber as an effective alternative to the typical free-space system due to its increased spectroscopic sensitivity[21]. Future work will focus on experimental improvements such as the use of polarization maintaining fiber, the integration of a Berek compensator and using a larger magnetic field strength. With these enhancements, this nanofiber system can be relatively easily implemented as a laser frequency stabilization method with definite advantages over the typical free-space configuration. For example, a high intensity is produced in the evanescent field for a low input power of just several nW and the mode area of the waist region is very small compared to a standard glass cell which, in turn, leads to enhanced light-matter interactions. Furthermore, we will also explore the miniaturization of the vacuum setup to reduce the system's footprint while improving the overall performance of this technique.

## ACKNOWLEDGEMENTS

This work was supported by Science Foundation Ireland under Grant No. 07/RFP/PHYF518s2 and OIST Graduate University. AW and JW acknowledge support from IRCSET through the Embark Initiative and the INSPIRE program, respectively.